\documentclass[aps,pra,reprint,amsmath,amssymb,superscriptaddress,longbibliography]{revtex4-2}
\usepackage{color,graphicx,float,hyperref,stackengine,braket,bm,mathtools,calc,float,eqnarray,subcaption,tikz,caption,ragged2e,soul}
\begin{document}

\title{From scale-free to Anderson localization: a size-dependent transition}

\author{Burcu \surname{Yılmaz}}
\affiliation{Department of Physics, Bilkent University, Ankara 06800, Türkiye}
\author{Cem \surname{Yuce}}
\affiliation{Department of Physics, Eskişehir Technical University, Eskişehir 26555, Türkiye}
\author{Ceyhun \surname{Bulutay}}
% \email{bulutay@fen.bilkent.edu.tr}
\affiliation{Department of Physics, Bilkent University, Ankara 06800, Türkiye}

\date{\today}
\begin{abstract}
Scale-free localization in non-Hermitian systems is a distinctive type of localization where the localization length of certain eigenstates, known as scale-free localized (SFL) states, scales proportionally with the system size. Unlike skin states, where the localization length is independent of the system size, SFL states maintain a spatial profile that remains invariant as the system size changes. We consider a model involving a single non-Hermitian impurity in an otherwise Hermitian one-dimensional lattice. Introducing disorder into this system transforms SFL states into Anderson-localized states. In contrast to the Hatano-Nelson model, where disorder typically leads to the localization of skin states and a size-independent Anderson transition, the scale-free localization in our model causes a size-dependent Anderson transition.
\end{abstract}

\maketitle

\section{Introduction}
Random potentials introduced throughout a periodical lattice cause scatterings and disruption of wave diffusion. As reasoned by Anderson, charge or spin transport behavior in disordered lattices can remarkably change from that of a periodic one having extended wave functions. The excitations can get trapped in the vicinity of impurities and cease to diffuse, causing their localization. This disorder-driven phenomenon is known as Anderson localization \cite{anderson1958}. 

In Hermitian systems, bulk states are generally insensitive to local impurities except for a few impurity-induced bound states \cite{guo2023}. However, when a local \textit{non-Hermitian} impurity is introduced, it leads to significant changes in the system’s spectral properties \cite{wang2019,kim2021,ghosh2022,zhang2023,tomasi2023,kochergin2024}. Specifically, the breaking of Parity-Time (PT) symmetry can result in imaginary parts of the eigenenergies accompanied by topological transitions \cite{ryu2020,tang2020,tang2021,tang2022,acharya2022,padhan2024}. Recent studies have uncovered several aspects of non-Hermitian systems such as the non-Hermitian skin effect \cite{yao2018,jiang2019},
and scale-free localization \cite{li2020, li2021, yokomizo2021}. The former displays high sensitivity to boundaries which causes the edge localization of eigenstates, mobility edge in the non-Hermitian Anderson localization \cite{jiang2019,kaili2023,noronha2022,luo2021,luo2022,wangwang2023,cai2021,cai2022,rafiulislam2022}, and it motivated the generalization of Brillouin zone under the open boundary conditions (OBC) in both Hermitian and non-Hermitian systems \cite{yang2020}. Lately, non-Hermitian skin effect has also been studied as induced by on-site dissipations \cite{yi2020non}, under generalized boundary conditions \cite{yuce2022}, as well as in the presence of a Kerr-type nonlinearity \cite{yuce2021,manda2024}. On the other hand, scale-free localization reveals the effects of introducing an impurity with non-Hermitian coupling to the system where the localization length depends on the system size, unlike the Anderson localization which is known to be size-independent \cite{li2023,guo2023,molignini2023}. The scaling rule of the scale-free localization length is shown to be determined by the geometry of the generalized Brillouin zone \cite{li2024}. Moreover, scale-free localization has been investigated from the viewpoints of hybridization with skin states \cite{fu2023}, and many-body interactions \cite{wang2023}. Very recently, it is experimentally demonstrated in electric circuit lattices having a non-Hermitian defect \cite{xie2024}.

In this work, our objective is to put forward a \textit{size-dependent} Anderson localization. For this purpose we introduce an impurity having a non-reciprocal (thus, non-Hermitian) coupling into a Hermitian system. We determine the critical disorder strength $W_c$, which is the value where the Anderson transition gets completed, and manifest that $W_c$ and hence Anderson localization are size dependent.

\section{Model}
\begin{figure}
\centering
\begin{subfigure}[ht]{.5\textwidth}
   \caption{}
    % \vspace{0.5cm}
\includegraphics[width=0.65\textwidth]{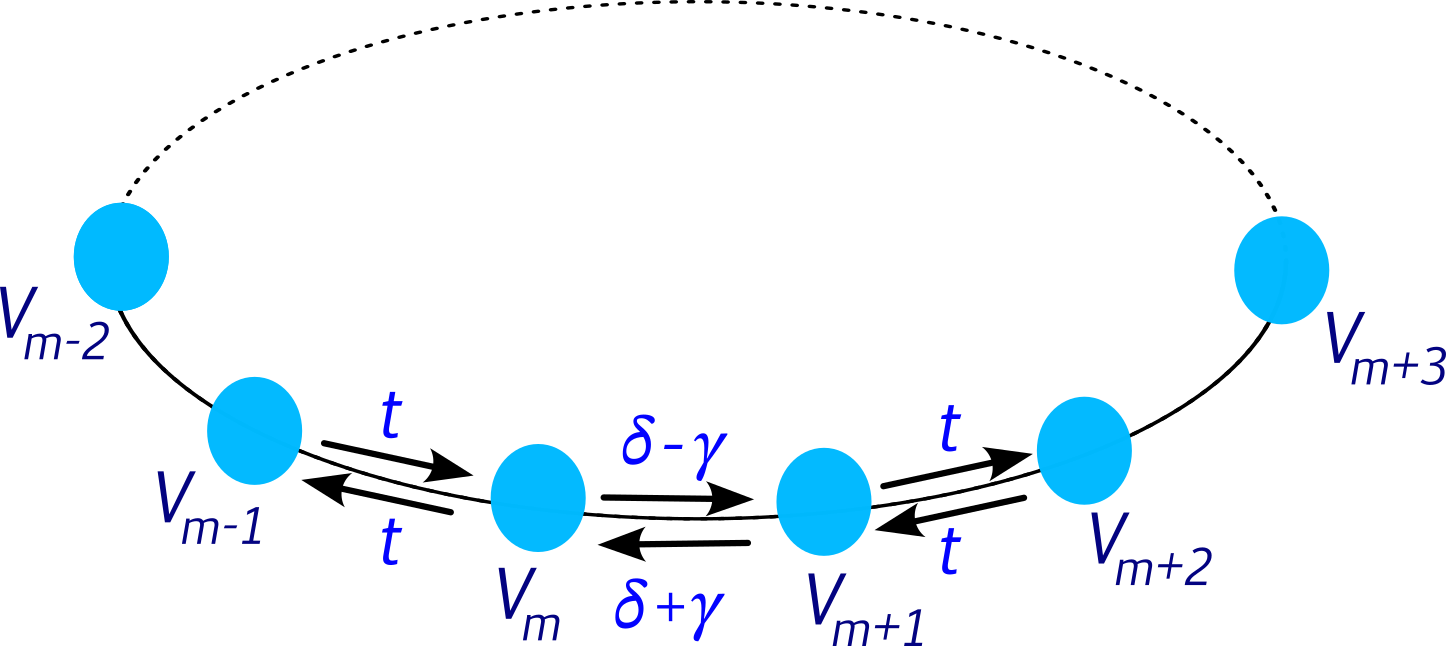}
    \label{fig:sketch2}
    % \vspace{0.5cm}
\end{subfigure}
% \hspace{0.05\textwidth}
\begin{subfigure}[ht]{.5\textwidth}
   \caption{}
\includegraphics[width=.9\linewidth]{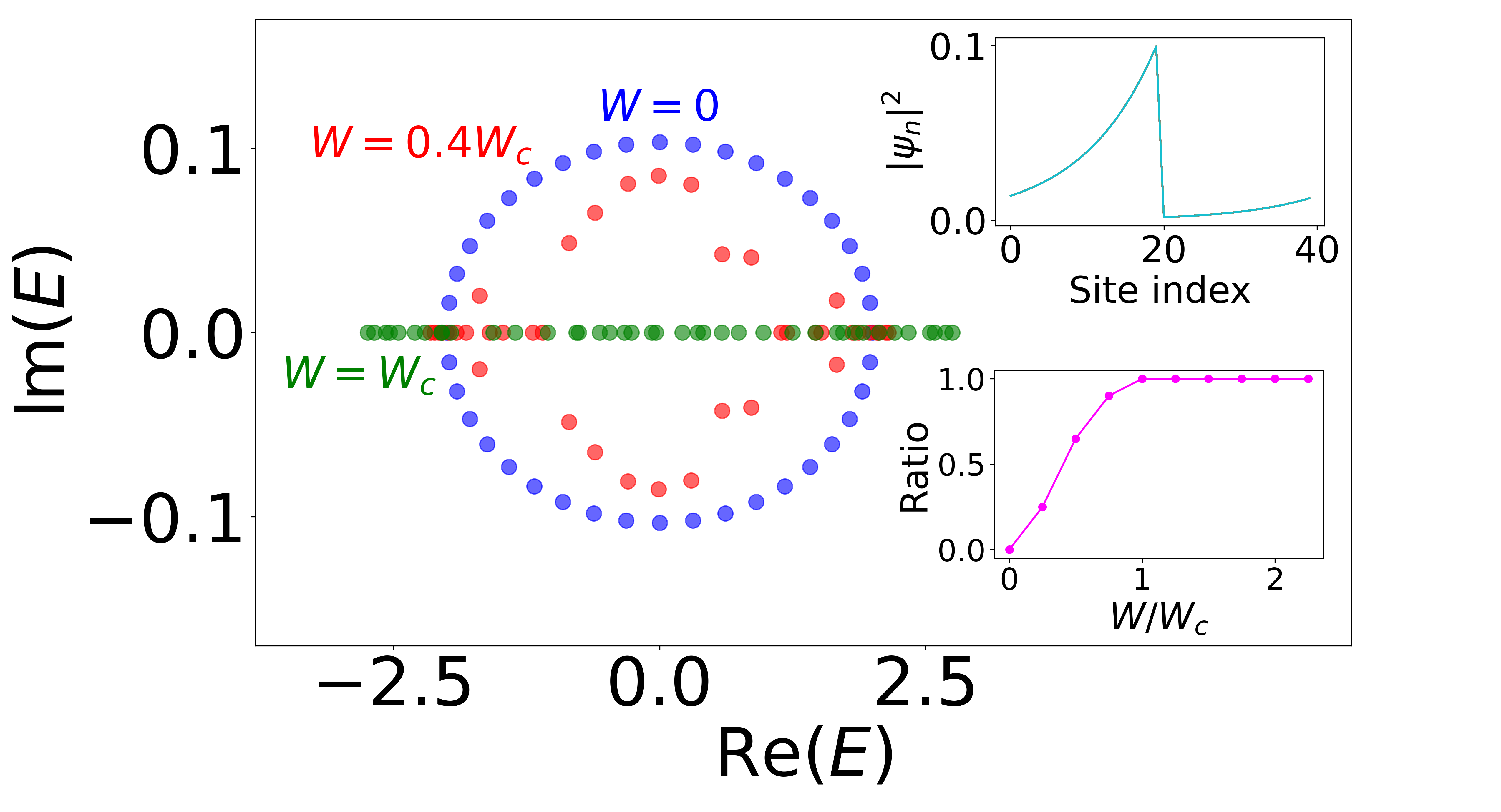}
    \label{fig:13-07-inset} 
\end{subfigure}
\caption{\justifying (a) 1D ring lattice of the Hermitian Anderson model with a non-Hermitian impurity. The Hermitian Anderson model shows an Anderson transition at any non-zero disorder strength. Adding a single impurity makes the Anderson transition point dependent on the system size. In the Hermitian Anderson model, extended states become Anderson-localized states at the Anderson transition point. However, when a single impurity is introduced, both extended and SFL states turn into Anderson-localized states at the transition point. This leads to a significant shift and size dependency in the Anderson transition point. (b) The PBC spectrum for three different values of $W$ at $\gamma=\sqrt{\delta^2  -1}$ for $N=40$. In the absence of the disorder, all states are SFL states, as illustrated in the above inset showing the probability density for all eigenstates. As $W$ increases, an increasing number of states are converted to Anderson-localized states with real energy. The lower inset shows the ratio of the Anderson-localized states over all eigenstates as a function of $W$. }
\label{fig:PBC-ipr}
\end{figure}

The Hermitian Anderson model exhibits an Anderson transition at any non-zero disorder strength. This behavior is observed for lattices of any size. However, the presence of a single non-Hermitian impurity can drastically modify this behavior, leading to a size-dependent Anderson transition. To study this phenomena, we consider a minimal model of a one-dimensional (1D) disordered Hermitian lattice with a non-Hermitian impurity, as illustrated in Fig.~\ref{fig:sketch2}. The Hamiltonian is given by
\begin{equation}
    \begin{aligned}
        \hat{H} &= \sum_{n \neq m} t(\hat{c}^{\dagger}_{n+1} \hat{c}_n + \text{H.c.}) 
        + \sum_{n=1}^{N} V_n \hat{c}^{\dagger}_{n} \hat{c}_n \\
        &\quad + (\delta - \gamma) \hat{c}^{\dagger}_{m+1} \hat{c}_{m} 
        + (\delta + \gamma) \hat{c}^{\dagger}_m \hat{c}_{m+1},
    \end{aligned}
    \label{eq:hamiltonian}
\end{equation}
where $t$ is the hopping constant between neighbouring sites, $\hat{c}_n$ $(\hat{c}^\dagger_n)$ annihilation(creation) operators at site $n$,  the real numbers $\delta$ and $\gamma$ define the local non-Hermitian impurity at site $m$, $V_n$ represents random uncorrelated on-site energies uniformly distributed in the interval of $W[-0.5, 0.5]$ with the real number $W$ being the disorder strength. Using the state, $\ket{\psi} = \sum_{n} \psi_n \hat{c}^{\dagger}_n \ket{0}$, the single particle Schrödinger equation $\hat{H}\ket{\psi} = E \ket{\psi}$ leads to the following discrete equations for the complex field amplitude $\psi_n$ at site 
$n \neq m, m+1$ 
\begin{equation}
    \begin{aligned}
    &t\psi_{n+1} + t\psi_{n-1} + V_n \psi_n = E \psi_n,\\
    \end{aligned}
    \label{eq:nnotm}
\end{equation}
and for the impurity sites at $n=m,m+1$ 
\begin{equation}
    \begin{aligned}
    &t\psi_{m-1} +(\delta + \gamma)\psi_{m+1} + V_m \psi_m = E \psi_{m},\\
    &t\psi_{m+2} +(\delta - \gamma)\psi_{m} + V_{m+1} \psi_{m+1} = E \psi_{m+1}.
    \end{aligned}
    \label{eq:n=m}
\end{equation}

Let us begin by briefly discussing the disorder-free system, $W = 0$ \cite{guo2023}. 
For simplicity, in the remainder of the treatment we set $t = 1$. Under periodic boundary conditions (PBC), the energy spectrum at $W=0$ may form a loop in the complex energy plane, whereas under OBC, it lies on the real axis. This spectral difference leads to scale-free localization under PBC, which is not present under OBC. The phenomenon of scale-free localization challenges conventional understanding of localization, as a single non-Hermitian impurity in an otherwise Hermitian lattice can generate a significant number of scale-free localized (SFL) modes. The SFL states that arise from a single coupling impurity in an otherwise Hermitian lattice can be analyzed using the PT-symmetry properties of the system; where P represents the parity, and T is the time-reversal operator \cite{guo2023}. By varying the parameters $\delta$ and $\gamma$, the system makes transitions between PT-unbroken, PT-broken, and PT-restoration regions. In the PT-unbroken region, where $ \gamma < |\delta - 1|$, the system has extended bulk states and two impurity-bound states with real eigenvalues under both PBC and OBC. When $\gamma >|\delta - 1|$, the system enters the PT-broken region, where the energy spectra begin to exhibit complex eigenvalues under PBC, leading to the formation of SFL states. The localization lengths of the SFL states vary proportionally with system size, implying that their spatial profiles remain invariant regardless of the system size. Note that extended and SFL states coexist in the system, with all eigenstates becoming SFL states at $\gamma=\sqrt{\delta^2 - 1}$. Further increasing $\gamma$ moves the system into the PT-restoration region, where $\gamma > \delta + 1$, which contains two bound states with complex eigenvalues and extended bulk states with real eigenvalues. 

Introducing random on-site energies into our system $(W \ne 0)$ converts extended and SFL states into Anderson-localized states. All the extended states transform into Anderson-localized states at a non-zero value of the disorder strength, similar to the Hermitian Anderson model. On the other hand, each SFL mode with a complex energy collapses towards the real-energy axis at its specific disorder strength, indicating that the fraction of SFL states decreases with increasing disorder strength. In other words, as $W$ increases, more SFL states coalesce on the real-energy axis and become Anderson-localized states, as shown in Fig.~\ref{fig:13-07-inset} at $\gamma=\sqrt{\delta^2 -1} $, where all states are SFL states in the absence of disorder. The critical disorder strength $W_c$ is identified by the value of $W$ at which the energy spectra of a lattice under both PBC and OBC are real valued, except for the energies of the two bound states. At this point, the system no longer exhibits high sensitivity to boundary conditions.

\begin{figure*}
\centering
\includegraphics[width=\textwidth]{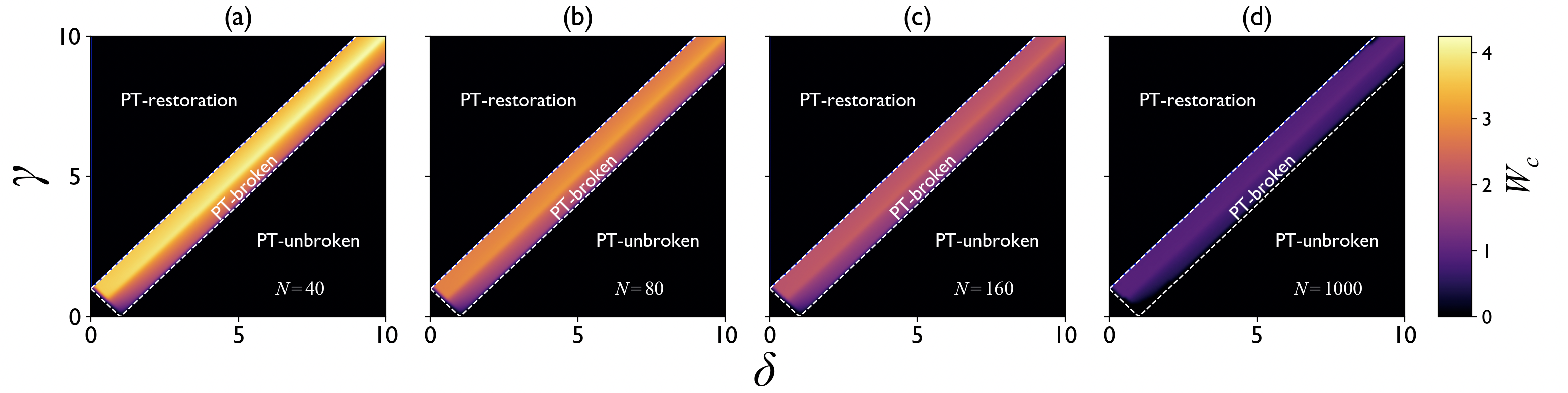}
\caption{\justifying Phase diagram of the critical disorder strength $(W_c)$ for different system sizes at $m=N/2$: (a) $N=40$, (b) $N=80$, (c) $N=160$, (d) $N=1000$ for $t=1$. White dashed lines are given by $\gamma = |\delta - 1|$ and $\gamma = \delta + 1$ and represent the phase boundaries of the PT-symmetry for there regions. PT-unbroken region where $ \gamma < |\delta - 1|$, PT-broken region where $|\delta - 1| < \gamma < \delta + 1$, PT-restoration region where $ \gamma > \delta + 1$. Results are averaged over many random potential realizations. Anderson and scale free localizations compete with each other, and SFL and Anderson-localized states coexist until the Anderson transition occurs at a critical value.}
\label{fig:phase_trans}
\end{figure*}

\section{Results}
We perform numerical calculations of Eqs.~(\ref{eq:nnotm}) and (\ref{eq:n=m}) to determine the critical disorder strength $W_c$ as a function of $\delta$ and $\gamma$ for various lattice sizes and present our results in Fig.~\ref{fig:phase_trans}.  
Since $W_c$ varies with each choice of random potentials, it naturally exhibits fluctuations from one realization to another. To account for this sample-to-sample variability, we average $W_c$ over many different realizations of the random potential. Below, we explore this in the context of Anderson localization in the PT-broken region, following a brief discussion of the PT-unbroken and PT-restoration regions.

In the PT-unbroken region, the energy spectra under PBC and OBC lie along the real axis in the complex energy plane, even in the absence of disorder. Like the Hermitian Anderson model, this system undergoes an Anderson transition by introducing infinitesimally small random potentials. Similarly, in the PT-restoration region, the energy spectra under both PBC and OBC also lie along the real axis, with the exception of two impurity-bound states that have complex eigenvalues in the absence of disorder. Excluding these two bound states induced by the non-Hermitian local impurity,  an infinitesimally small random potential leads to the localization of eigenstates. 

In the PT-broken region, the energy spectrum exhibits high sensitivity to boundary conditions, meaning it can be significantly altered by introducing boundaries. This spectral difference between PBC and OBC implies distinct characteristics for the corresponding eigenstates, with extended and SFL states appearing under PBC. Disorder suppresses boundary sensitivity and drives the PBC spectrum toward real-valued energies. In other words, the extended and SFL states begin to transform into localized ones, with the extended states transitioning first, followed by the SFL states. As the disorder strength increases gradually, the eigenstates are sequentially converted into Anderson-localized states, with the SFL state that has the highest imaginary part of the energy being the last one to become Anderson localized. Once the final SFL state with complex energy transitions to real energy, the system no longer exhibits sensitivity to boundary conditions, indicating fully Anderson-localized behavior. The overall process represents a progressive crossover rather than a sharp phase transition.

\begin{figure}
\centering
\includegraphics[width=0.7\linewidth]{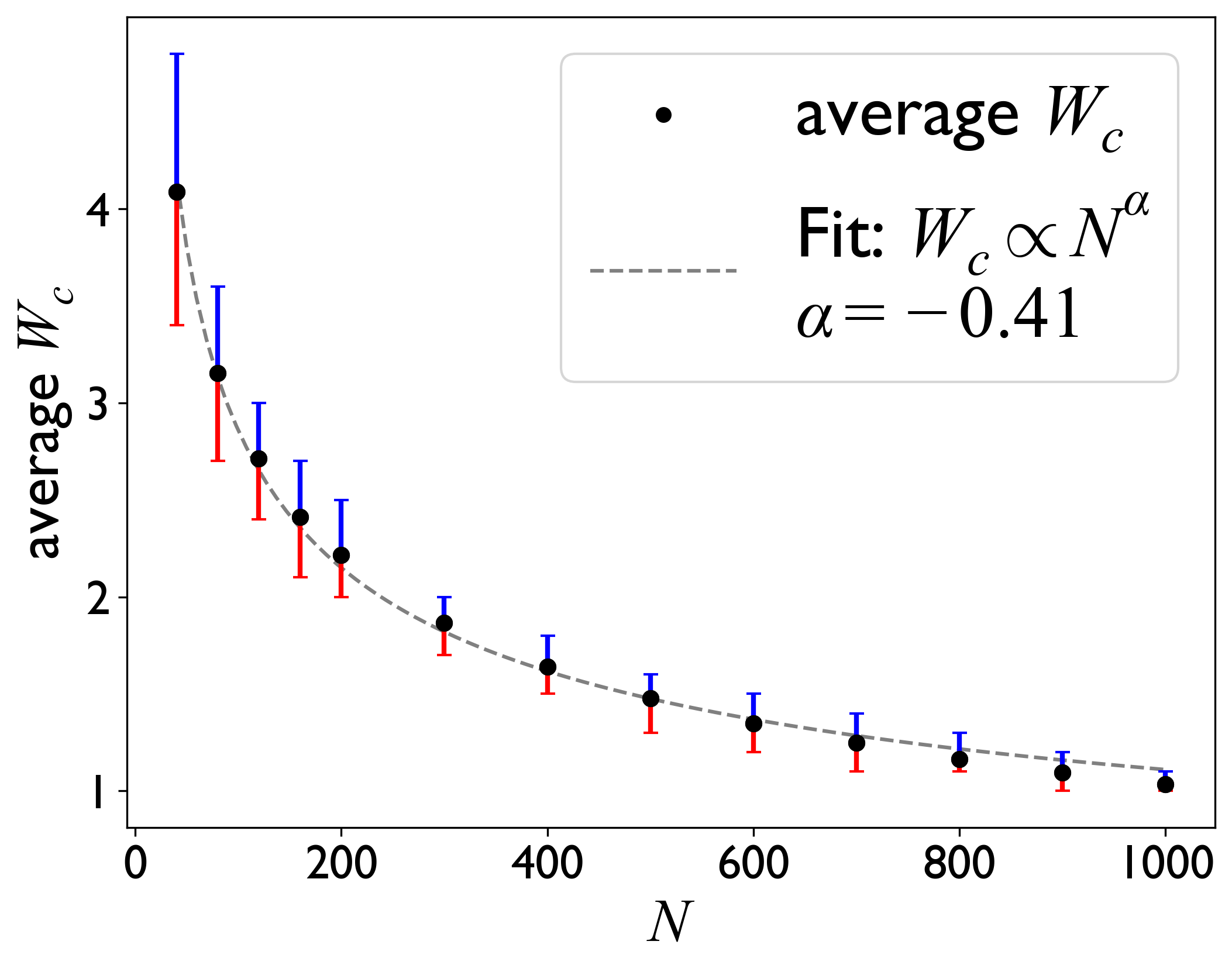}
\caption{\justifying Scaling relationship between the critical disorder strength $W_c$ and the lattice size $N$. Black dots represent the average critical disorder strength calculated for various lattice sizes, while the gray dashed line indicates the best fit to the scaling law $W_c \propto N^{\alpha}$, where $\alpha = -0.41$ with $\gamma = \delta = 6.5$. Asymmetrical error bars indicate the sample standard deviations on either side of the mean value, calculated over 10\,000 realizations.}
\label{fig:scaling-relationship}
\end{figure}

As another useful remark we note that the fraction of extended states to SFL states at $W=0$ varies with the parameters $\delta$ and $\gamma$. Therefore, the Anderson transition point changes with these parameters. The critical disorder strength  $W_c$ marks the transition where extended and SFL states become Anderson-localized states, consequently removing the high sensitivity of the spectrum to the boundary conditions. $W_c$ takes its maximum value at $\gamma = \delta $ as can be seen in Fig.~\ref{fig:phase_trans}. Moreover, there is a sharp transition in $W_c$ at the boundary between the PT-broken and other regions. In the PT-broken region with a fixed $\delta$, $W_c$ increases until it reaches its maximum value at $\gamma = \delta $ where $\delta \geq \frac{1}{2}$, this behavior can be attributed to the semi-edge effect that emerges at $\gamma = \delta$, where a non-Hermitian impurity in the bulk creates a boundary-like behavior by having zero coupling in one direction and non-zero coupling in the other. This asymmetry enhances localization as the impurity effectively acts as a boundary in one direction. $W_c$ then decreases until it reaches the border between the PT-broken and PT-restoration regions. This shows how a non-Hermitian coupling impurity in the Hermitian lattice changes the Anderson transition point, a phenomenon not occurred in Hermitian systems, as a single Hermitian impurity does not affect the Anderson localization transition. 
Figure~\ref{fig:phase_trans} also reveals size-dependency of the Anderson transition: namely, increasing the lattice size from $N = 40$ to $1000$ leads to a decrease in the critical disorder strength, $W_c$. As $N$ increases, Anderson transition occurs at weaker disorder strength, as the effect of a single impurity diminishes with larger lattice sizes. Note that $W_c$ approaches zero in the thermodynamic limit. 

Figure~\ref{fig:scaling-relationship} illustrates the scaling relationship between the critical disorder strength, $W_c$, and the system size, $N$. By fitting our numerical data, indicated by the gray dashed line, we find that $W_c \propto N^{\alpha}$, where $\alpha$ is the scaling exponent. The value of $\alpha$ is found to vary between $-0.36$  and $-0.41$ depending on the parameters $\gamma$ and $\delta$ when averaged over 10\,000 realizations. The negative exponent \( \alpha \) suggests that the critical disorder strength decreases as the system size increases, highlighting the inverse relationship between \( W_c \) and \( N \). Unlike the standard thermodynamic phase transitions, $W_c$ displays variations among different realizations of random disorder. To depict the range of variation in $W_c$ for each system size $N$, error bars are included in Fig.~\ref{fig:scaling-relationship} which specify one standard deviation (34\% confidence interval) on either side of the mean values. This reveals that the distribution in $W_c$ has a positive skewness as $N$ increases. Concomitantly, as expected the error bars rapidly shrink as $N$ increases.

In the PT-unbroken and PT-restoration regions, Anderson transition occurs at arbitrary values of the disorder strength. However, a lattice has a finite length in practice, and the localization lengths of some eigenstates may exceed the system size even if $W$ is greater than $W_c$. Consequently, the corresponding states behave as if they are delocalized. Similarly,  Anderson-localized states may not be observed if the localization lengths of the eigenstates can surpass the system size in the PT-broken region. To study the localization properties of the eigenstates, we use the inverse participation ratio (IPR) as a measure for the localization of a specific eigenstate with energy $E$. It is defined as
\begin{equation}
    \begin{aligned}
    \mathrm{IPR} = \frac{\sum\limits_{n} \left| \psi_n \right|^4}{\left( \sum\limits_{n} \left| \psi_n \right|^2 \right)^2}
    \end{aligned}
\end{equation}
where $\psi_n$ represents the eigenstate at energy $E$. A high IPR value close to 1 indicates that the state is highly localized. On the other hand, a low IPR value indicates that the state is delocalized. For a completely delocalized state where all components have equal amplitude, the IPR approaches $1/N$. Each state exhibits distinct IPR behaviors before and after the occurrence Anderson transition. We also define the average IPR value as the average of all individual IPRs.

\begin{figure}
\centering
\begin{subfigure}[b]{.45\textwidth}
   \centering
    \includegraphics[width=.85\textwidth]{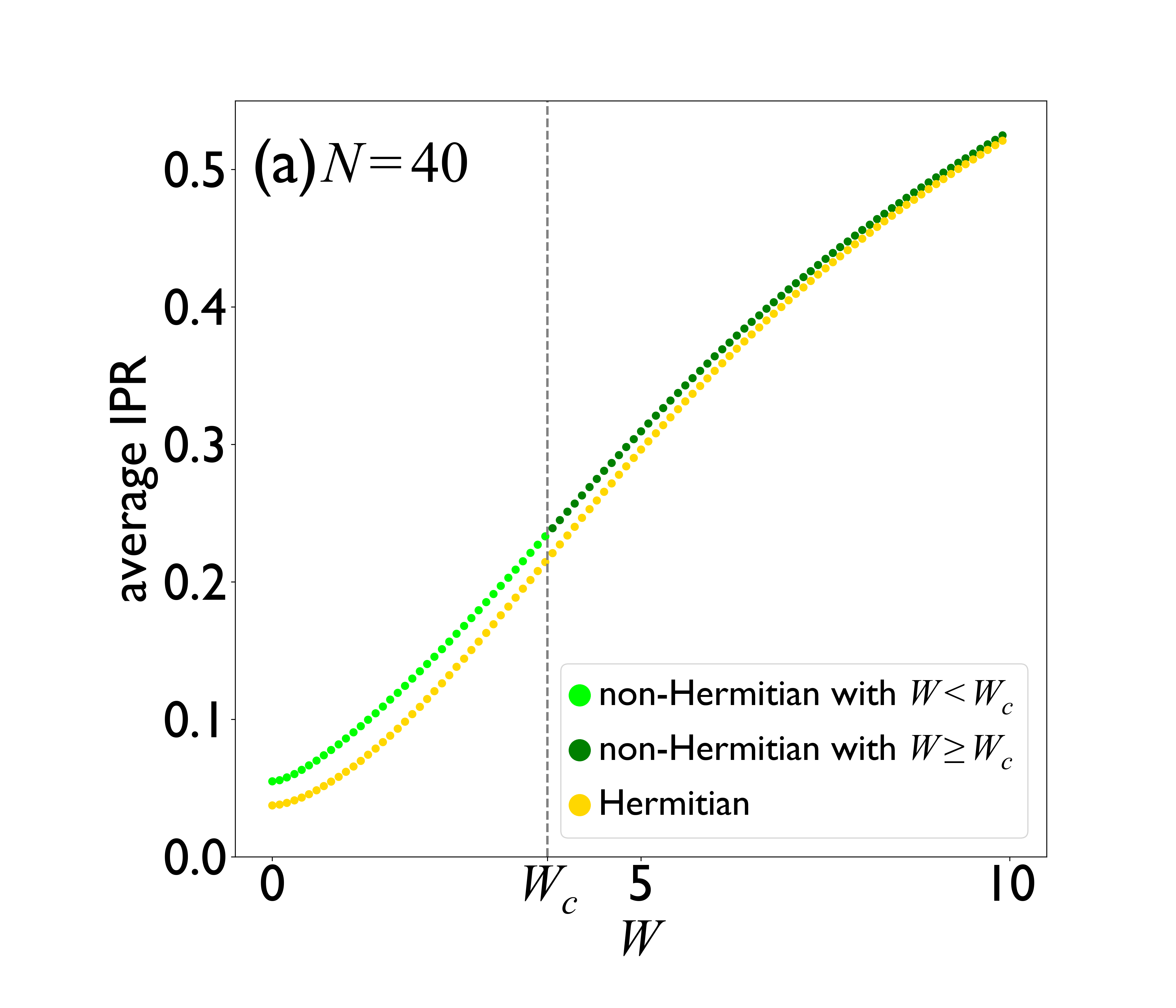}
    \label{fig:N=40-min(IPR)}
\end{subfigure}
\begin{subfigure}[b]{.45\textwidth}
\centering
\includegraphics[width=.85\linewidth]{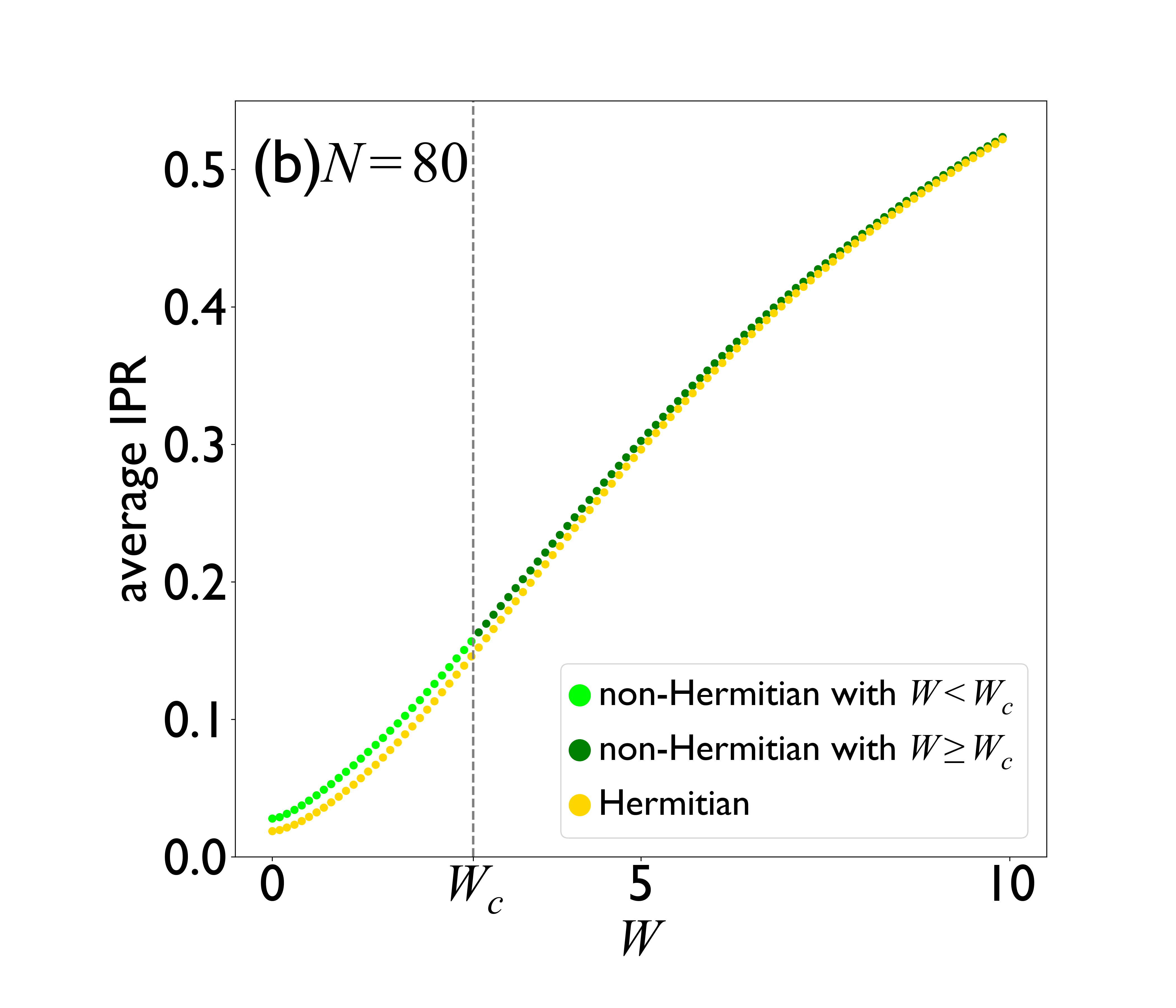}
    \label{fig:N=80-min(IPR)} 
\end{subfigure}
\caption{\justifying The average IPR as a function of disorder strength $(W)$, averaged over 10\,000 realizations for both Hermitian Anderson model ($\gamma =0, ~\delta = 1$) and non-Hermitian Anderson model when $\gamma=\delta=4$ for (a) $N=40$ and (b) $N=80$.}
\label{fig:min(IPR)}
\end{figure}

\begin{figure*}[t]
    \centering
    \includegraphics[width=\textwidth]{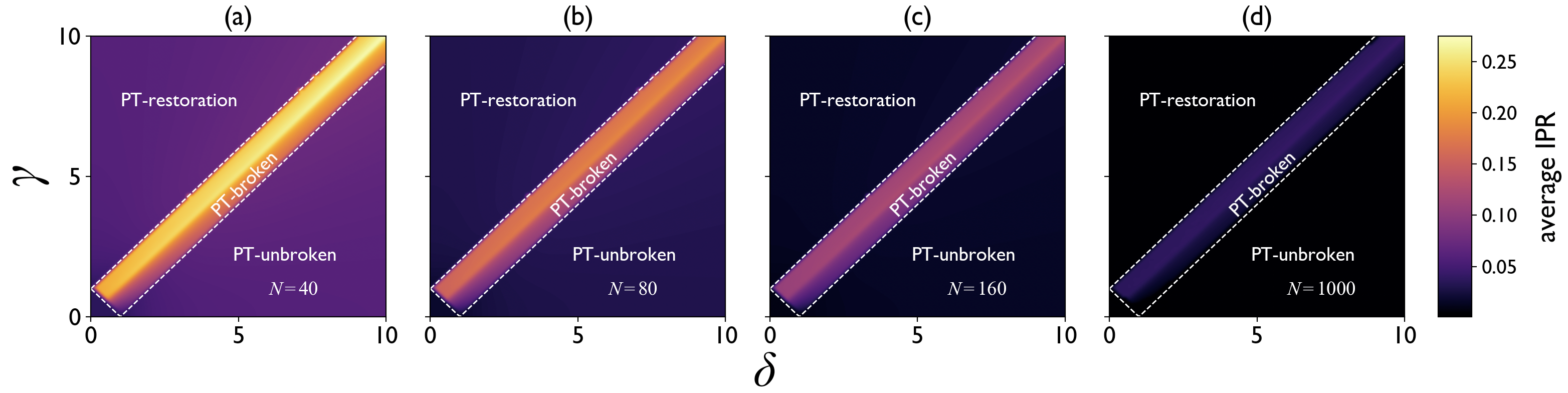}
    \caption{\justifying Average IPR values at \(W = W_c\) for various system sizes at $m=N/2$: (a) \(N = 40\), (b) \(N = 80\), (c) \(N = 160\), and (d) \(N = 1000\). The white dashed lines are given by \(\gamma=|\delta - 1|\) and \(\gamma = \delta + 1\). Notice the difference in average IPR values between the PT-broken region and the other two regions at the same disorder strength. This difference arises because SFL states transform into Anderson-localized states in the PT-broken region, whereas extended states undergo this transformation in the other regions.}
    \label{fig:ipr-values}
\end{figure*}

Figure~\ref{fig:min(IPR)} presents the average IPR value for our model, averaged over 10\,000 realizations, and compares them to those of the Hermitian Anderson model. As the disorder strength $W$ increases, the average IPR values also increase, indicating that the eigenstates are becoming more localized. In the non-Hermitian case, the rate of IPR increase is high up to the critical disorder strength $W_c$, which indicates the ongoing SFL to Anderson-localized conversion. Beyond $W_c$, the increase in average IPR values slows down, and the behavior for the non-Hermitian case start to overlap with the Hermitian model, justifying that both systems possess Anderson-localized states. This overlap behavior is more visible for the $N=80$ case. It is noteworthy that the size dependency of $W_c$ can also be seen in this figure by comparing $N=40$ with $80$.

Let us now examine average IPR values at $W_c$  shown in Fig.~\ref{fig:ipr-values} for various $N$, averaged over many realizations. At the critical disorder strength, the average IPR values are moderate, suggesting that a large portion of eigenstates are localized with smaller localization lengths than the system size, while a few are acting as if they are delocalized as their localization lengths exceed the system size, like those in Hermitian Anderson model. As $N$ increases, the average IPR values decrease and approach zero in the thermodynamic limit, indicating that the size dependency is more pronounced for smaller $N$ values. It is interesting to note that the average IPR values exhibit a sharp transition at the boundaries between the PT-broken regions and other regions at the same disorder strength. In other words, the Anderson-localized states have shorter localization lengths in the PT-broken regions than in the other regions at same $W$. This is due to the interplay between two types of localization: scale-free localization and Anderson localization. The relationship between localization length and disorder strength in the PT-unbroken region can be approximated similarly to that 
in a Hermitian system, where it is inversely proportional to the square of the disorder strength. Conversely, in the PT-broken region, 
the localization length is inversely proportional to the system size in the absence of disorder, and increases with disorder, as illustrated 
in Fig.~\ref{fig:ipr-values}. Here, it can be observed that when all SFL states transition into Anderson-localized states at the critical disorder strength, the inverse proportionality to disorder strength changes from super-linear to sub-linear dependence on $W$.

\section{Discussion}
We should emphasize the subtle differences between our model and similar ones in the literature. Notably, our model differs from the Hatano-Nelson model, where disorder induces a transition from skin states to Anderson-localized states, where skin states are characterized by size-independent localization lengths \cite{hatano1996,gong2018,zeng2020,liu2021exact,sen2022,han2022,tang2024}. Consequently, in the Hatano-Nelson model, the Anderson localization transition occurs at a non-zero disorder strength and is independent of system size. Additionally, our model is different from the Hermitian Anderson model, in which all states are extended in the absence of disorder but localize when disorder is introduced. In this Hermitian Anderson model, the Anderson transition point is zero and independent of system size. In contrast, our model shows that both extended and  SFL states transition to Anderson-localized states, leading to a size-dependent, non-zero transition point due to the size-dependent nature of SFL states.

For a wider perspective, SFL states may coexist with size-independent localized states (such as skin states) or exist within a system that has both complex OBC and PBC spectra, exhibiting significant differences. Introducing onsite disorder deforms both spectra until the disorder strength reaches a critical level, at which point the boundary condition sensitivity is lost and both the OBC and PBC spectra coincide in the complex energy plane \cite{gong2018}.

\section{Conclusion}
To summarize, we have studied the size-dependent Anderson localization transition in a non-Hermitian system. In the Hermitian Anderson model, extended states become Anderson-localized states upon the introduction of infinitesimally small disorder, indicating a zero critical disorder strength. Conversely, in the Hatano-Nelson model, skin states transform into Anderson-localized states when disorder is introduced, resulting in a finite critical disorder strength. In a Hermitian Anderson lattice with a single non-Hermitian coupling impurity, SFL states are converted into Anderson-localized states in the presence of the disorder. Given that SFL states have localization lengths that scale with the system size, their conversion to Anderson-localized states becomes size-dependent. This leads to a size-dependent critical disorder strength for the Anderson transition. As the lattice size increases, this critical disorder strength decreases and approaches zero in the thermodynamic limit. This implies that larger systems require less disorder to achieve Anderson localization, with $W_c$ approaching zero in the thermodynamic limit.

\section*{Acknowledgement}
The numerical calculations reported in this paper were partially performed at
T\"UB\.ITAK ULAKB\.IM, High Performance and Grid Computing Center (TRUBA
resources).
% \section*{Appendix}

%

\end{document}